# Interconnection between 802.15.4 Devices and IPv6: Implications and Existing Approaches

Md. Sakhawat Hossen[1#], A. F. M. Sultanul Kabir[2], Razib Hayat Khan[3] and Abdullah Azfar[1*]

[1] School of Information and Communication Technology,
Royal Institute of Technology (KTH), Stockholm, Sweden

[2] Department of Computer Science,
American International University of Bangladesh (AIUB), Dhaka, Bangladesh

[3] Department of Telematics,
Norwegian University of Science and Technology (NTNU), Trondheim, Norway

**Abstract**
The increasing role of home automation in routine life and the rising demand for sensor networks enhanced wireless personal area networks (WPANs) development, pervasiveness of wireless & wired network, and research. Soon arose the need of implementing the Internet Protocol in these devices in order to WPAN standards, raising the way for questions on how to provide seamless communication between wired and wireless technologies. After a quick overview of the Low-rate WPAN standard (IEEE 802.15.4) and the Zigbee stack, this paper focuses on understanding the implications when interconnecting low powered IEEE 802.15.4 devices and a wired IPv6 domain. Subsequently the focus will be on existing approaches to connect LoWPAN devices to the internet and on how these approaches try to solve these challenges, concluding with a critical analysis of interoperability problems.

## 1. Introduction

The Institute of Electrical and Electronics Engineers (IEEE), an international non-profit organization for the advancement of technology related to electricity, is a major international standards body, with nearly 900 active standards. The IEEE 802 Local and Metropolitan Area Network Standards Committee (LMSC) of IEEE aims to develop standards for Local Area Networks (LAN) and Metropolitan Area Networks (MAN) carrying variable-size packets, and to specify services and protocols of the lower two layers (data link and physical) of the seven-layer OSI networking reference model. It also splits the OSI Data Link Layer in two sub-layers: Logical Link Control (LLC) and Media Access Control (MAC). The most famous IEEE 802 families are the 802.2 LLC, 802.3 Ethernet, 802.5. Token Ring, 802.11 Wireless LAN, 802.16 Broadband Wireless, and 802.15 Wireless PAN [1].

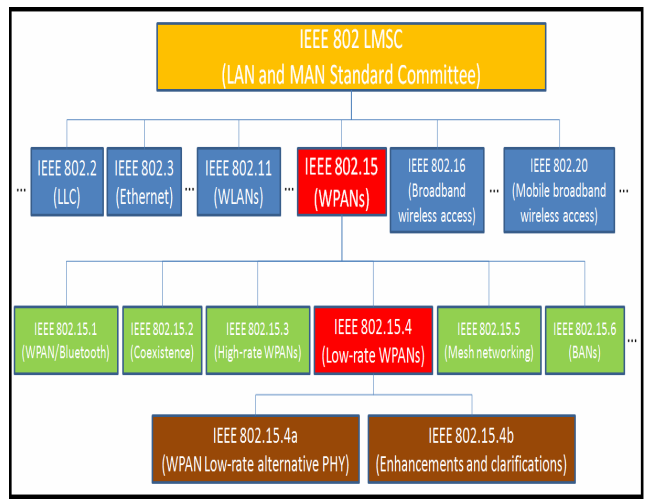

Fig. 1 IEEE 802 main standards

A Personal Area Network (PAN) is a network for interconnecting devices centered on an individual person's workspace, typically a short range. Personal area networks may be wired with computer buses (USB, FireWire) or wireless, and in this case they are called WPAN. IEEE 802.15 is the standard for Personal Area Networks or short distance wireless networks. These WPANs address wireless networking of portable and mobile computing devices such as PCs, PDAs, peripherals, cell phones, and consumer electronics. IEEE 802.15 includes different task groups, such as Bluetooth (802.15.1), the High-Rate WPAN (802.15.3), and the Low-Rate WPAN (802.15.4). [4]

This paper is focused on the Low-Rate WPANs, and it is organized as follows. Section 2 to 4 give an overview of IEEE 802.15.4 and Zigbee standards, for a better understanding of the following part of the paper. Section 5 analyzes the needs and the challenges of LoWPANs' connection to IPv6, whereas section 6 carefully analyzes existing approaches and compares them. The final part of the paper is dedicated to interoperability considerations.

## 2. IEEE 802.15.4 and Zigbee

Low-Rate WPANs are characterized by very low duty cycle, very long primary battery life (duration from months to years), low cost, and support for large number of nodes. Ease of design and deployment are key characteristics of these devices, which have the ability to remain quiescent for long periods of time without communicating, this is important due to power saving constraints. It is completely different from Bluetooth, the IEEE 802.15.1 standard for WPANs that is characterized by a higher data rate and objectives including handling voice, images, and file transfers in *ad hoc* networks. Moreover, the latency time to wake up when snoozing of low-rate WPANs is around 15 ms, instead of the 3 seconds for Bluetooth. The data rate of LoWPANs goes from 20 kbps to 250 kbps, depending on the transmitting frequency (shown in Table 1), and this is an indicator that those kinds of devices are not designed for video streaming or big file transfers.

IEEE 802.15.4 is the IEEE standard for low data rate wireless PANs, which focuses on the specification of the two lower layers of the protocol: physical and data link layers. It does not provide the upper layers of the protocol stack. Zigbee technology is a low data rate, low power consumption, and low cost wireless networking protocol targeted towards automation and remote control applications. IEEE joined forces with the Zigbee Alliance and worked closely to specify the entire protocol stack: IEEE 802.15.4 focused on the lower two layers of the protocol stack, whereas Zigbee Alliance aims to provide the upper layers for interoperable data networking, security services and a range of wireless home and building control solutions, provide interoperability compliance testing, marketing of the standard, advanced engineering for the evolution of the standard [2], [3], [4].

## 3. IEEE 802.15.4 Specification

3.1 The physical layer of IEEE 802.15.4

The PHY layer defines the physical and electrical characteristics of the network, for example, specifying the receiver sensitivity and transmitting output (in order to conform to national regulations) power. The basic task of this layer is thus data transmission and reception, and at the physical/electrical level, this involves modulation and spreading techniques that map bits of information in such a way as to allow them to travel through the air. The PHY tasks can be summarized as follows [5]:
• Enable/disable the radio transceiver (since low duty cycle saves energy);
• Compute Link Quality Indication (LQI) for received packets;
• Energy Detection (ED) within the current channel by means of signal strengths estimation;
• Listen to channels and declare availability or not (also called CCA - Clear Channel Assessment). There are three modes: Energy above threshold, Carrier sense only, Carrier sense with energy above threshold.

The physical layer services can be accessed through the Physical Data Service Access Point (PD-SAP) and the Physical Layer Management Entity Service Access Point PLME-SAP: PD-SAP provides data services (primitives PD-DATA.request, PC-DATA.confirm and PC-DATA.indication), whereas PLME-SAP provides the PAN Information Base Management Primitives (PLME-GET and PLME-SET to request and confirm, PLME-SET-TRX-STATE to enable and disable the physical interface, PLME-CCA for Clear Channel Assessment, and PLME-ED for energy detection) [6].

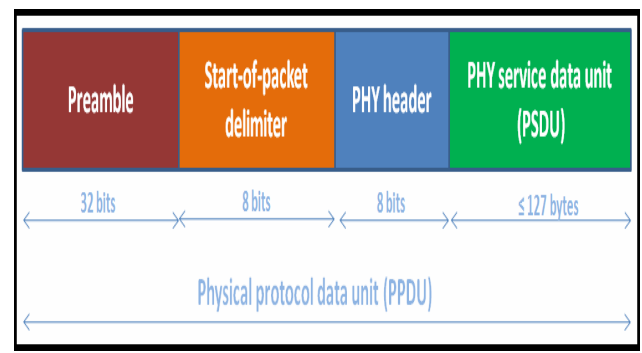

Fig. 2 IEEE 802.15.4 PPDU format

Figure 2 shows the Physical Protocol Data Unit (PPDU) frame, which is composed of a synchronization header consisting of a 4 byte preamble of binary zeros (for chip and symbol synchronization) and 1 octet for start-of-packet delimiter (SFD) that signifies end of preamble (0xE6 = '11100101'); 1 byte header (7 bits for the frame length plus 1 bit reserved for future use), and the payload (that must be less than 127 bytes). The PPDU can contain either data or data acknowledgment and its packet size varies from 5 to 127 bytes.

The physical layer specifies also the raw data rate characteristics, which can either be selected to offer a larger coverage area, or a higher throughput. In particular low rate of the 868/915 MHz physical layer can be translated into better sensitivity and a larger coverage area, thus reducing the number of nodes in a given area. While operation in the 2.4 GHz band can be used to attain higher throughput and lower latency or lower duty cycles.

It must be said that no license is required then 868 MHz is used in Europe or the 915 MHz band for Americas, whereas 2.4 GHz is available worldwide. Table 1 summarizes these all the details.

Table 1: 802.15.4 Physical layer characteristics

| Frequency band | Geographic region | Channel number(s) | License required? | Data parameters ||| Spreading parameters ||
|---|---|---|---|---|---|---|---|---|
| | | | | Bit rate | Symbol rate | Modulation | Chip rate | Modulation |
| 868.0 - 868.6 MHz | Europe | 0 | No | 20 | 20 Kbaud | BPSK | 0,3 Mchips/s | BPSK |
| 902.0 - 928 MHz | America | 1-10 | No | 40 | 40 Kbaud | BPSK | 0,6 Mchips/s | BPSK |
| 2.4 - 2.4835 GHz | Worldwide | 11-26 | No | 250 | 62,5 Kbaud | 16-ary orthogonal | 2,0 Mchips/s | O-QPSK |

### 3.2 The MAC layer of IEEE 802.15.4

The IEEE 802.15.4 MAC Standard provides information about type and association of devices, channel access mechanism, packet delivery, frame structure, guaranteed packet delivery, possible network topologies and security issues. In order to communicate with the upper layers it provides the MAC data service (MAC Common Part Sublayer, or MCPS-SAP) and the MAC management service (MLME-SAP). The MAC data service enables transmission of MAC protocol data units (MPDU) across the Physical Layer data service. The MAC sub layer features include beacon management, channel access, GTS management, frame validation, acknowledged frame delivery, association and disassociation. The MAC also provides support for implementing defined security mechanisms like AES-128, ACL modes, Data Encryption, Frame Integrity and Sequential Freshness.

In general the Logical Link Control sub-layer sits in top of the MAC layer, providing multiplexing of protocols transmitted over the MAC layer, optional flow control, and any retransmission of dropped packets. However, the IEEE 802.15.4 standard has modified and defined Layer 2 to either allow an IEEE 802.2 LLC to access the 802.15.4 MAC Sublayer through a Service Specific Convergence Sublayer (SSCS) as defined in Annex A of the IEEE 802.15.4 Standard or to allow direct access to the MAC by the upper layers, as used by proprietary networks and ZigBee networks [7].

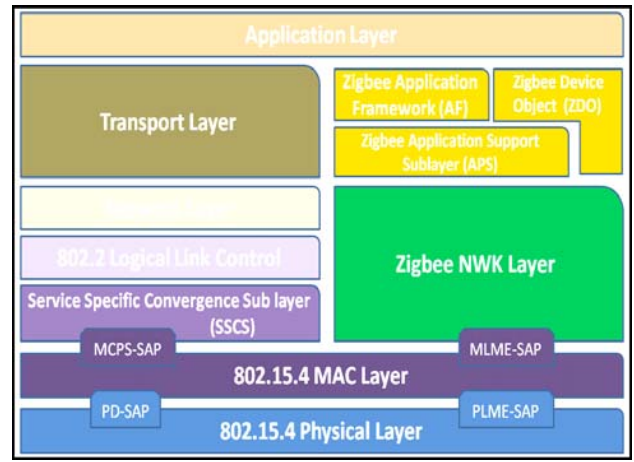

Fig. 3 Overview of the OSI layers with IEEE 802.15.4 and Zigbee standards

The Service Specific Convergence Sub layer (SSCS), as illustrated in figure 3, exists conceptually above the MAC Common Part Sub layer (MCPS).

IEEE 802.15.4 supports both short (16 bits) and extended (64 bits) addressing. An extended address (also called EUI-64) is assigned to every RFD that complies with the 802.15.4 specification. This means that a network can have up to 264 nodes. Moreover, when a device associates with a WPAN, it can receive a 16-bit address from its parent node that is unique in that network and is called the PAN ID. Each WPAN has thus a 16-bit number (PAN ID) that is used as a network identifier and it is assigned when the PAN coordinator creates the network. A device can try and join any network or it can limit itself to a network with a particular PAN ID [5].

In addition to the beacon frame, IEEE 802.15.4 also defines a data frame, an acknowledgment frame to confirm successful frame reception, and a command frame, used for handling all MAC peer entity control transfers. All four frames are characterized by header, payload, and trailer, but with different formats for each case. In particular, the ACK frame has not addressing fields and payload, so it's only 5 bytes, and is the minimal size MAC frame.

## 4. Zigbee Stack

Although sometimes Zigbee is considered to be the same as 802.15.4, it is conceptually completely different. The ZigBee protocol stack has its origins in the Open Systems Interconnect (OSI) seven-layer model, initiated in the early 1980s by ISO and ITU-T7. The Zigbee Stack has been developed by the Zigbee Alliance and it sits on top of the two layers defined by the IEEE 802.15.4 standard, which has been described in section 3. This protocol stack defines the two additional layers, the Zigbee Network Layer (NWK) and the Zigbee Application Layer, that are described in detail in this section. On top of this stack there are the application profiles to be followed by developers whenever they build devices. Each application profile has a unique profile identifier. The purpose of a profile is to create an interoperable, distributed application layer for separate devices. Profiles are simply standard rules and regulations [3]. There are three types of profiles: public (managed by the Zigbee Alliance), private (defined by an OEM (Original Equipment Manufacturer), and published (if an OEM decides to publish its private profile) [16].

A Zigbee stack provides all the functionality required by the Zigbee specification so that manufacturers can focus on developing their product's applications. If the manufacturer uses one of the public profiles, most of the configuration is already done. If none of the public profiles fits the manufacturer's needs, a new profile can be created, which can take advantage of the job already done by other profiles. A full ZigBee protocol stack is illustrated in [3].

Mobility (ability to move to different networks while maintaining communications) is fundamental as well as multihoming (capability for a node to be connected to multiple networks). Since NAT is not needed anymore, extra costs can be avoided, and this is important for cost prohibitive wireless sensor networks. If, in addition to IEEE 802.15.4, LOWPANs devices use other kind of network interfaces such as Ethernet or IEEE 802.11, the goal is to seamlessly integrate the networks built over those different technologies: this is a primary motivation to use IP to begin with. Despite all the challenges IPv6 over LoWPANs devices can represent a very appealing choice. For example, doing so enables the possibility of adding innovative techniques such as location aware addressing. [8], [9] and [21].

## 5. Existing Interconnection Challenges

LoWPANs intrinsic characteristics pone several challenges when designing an Internetworking solution. In this chapter we try to list the most important challenges [21]. These challenges will be examined again when describing existing approaches and possible solutions.

5.1 Frame size

The most evident problem of LoWPANs is the small packet size. In section 3, frame formats of IEEE 802.15.4 physical and MAC layer were shown. Given that the MAC layer maximum packet size is 127 bytes (133 bytes of the physical PDU minus the 6 bytes given by preamble, start-of-packet delimiter and PHY header), and 25 bytes are needed for Frame control, sequence number, addressing fields and FCS, only 102 octets are available in the MAC payload. But since Link-layer security imposes further overhead, which in the maximum case (21 octets of overhead in the AES-CCM-128 case, versus 9 and 13 for AES-CCM-32 and AES-CCM-64, respectively), the total number of bytes left available for data packets is 81.

Even taking into account that applications typically send small amounts of data, we must take into account that bulk data transfer may happen. If we add to those constraints the fact that the IPv6 header is 40 bytes long, TCP header is 20 octets and UDP header is 8 octets, the available space for data payload is furthermore reduced and very few bytes are left for data.

5.2 Fragmentation issues

Since applications do not know the constraints of physical links that might carry their packets, we should make sure that interoperability is possible. For this reason fragmentation is a key issue, especially in function of the fact that IP packets are very large compared to 802.15.4 maximum frame size; in fact IPv6 requires all links to support 1280 byte packets [22].

5.3 LoWPANs design constraints

Even if fragmentation is allowed, applications within LoWPANs are expected to originate small packets. Adding all layers for IP connectivity should still allow transmission in one frame, without incurring excessive fragmentation and reassembly. But since IPv6 has requirements of sub-IP reassembly, we have a challenge that comes from the intrinsic constraints of LoWPANs devices, which are simple devices, with low resources and demanding cost, power and energy saving constraints. In fact, RFDs may not have enough processing, RAM, or storage for a 1280 byte packet.

Another intrinsic behavior of LoWPANs is sleeping for long periods of time in order to conserve energy: this means devices are unable to communicate during these sleep periods. Unreliability due to radio connectivity, battery drain, and device lockups must be taken into account.

Concerning address space, a large number of devices are expected to be deployed in the long run; however this does not scare IPv6 addressing features.

### 5.4 LoWPAN's addressing

Dealing with the two different addressing methodologies available in 802.15.4 must be clearly taken into account, as well as how to perform stateless auto configuration (as compared to stateful), because of the reduction of the configuration overhead on the hosts. A method to generate an "interface identifier" from the EUI-64 assigned to the IEEE 802.15.4 device is thus needed.

### 5.5 Service discovery

Current service discovery methods are "heavyweight" for LoWPANs: they are primarily heavyweight protocols based on XML such as SOAP, thus not suitable for LoWPANs, which require simple service discovery network protocols to find, control, and maintain services provided by devices. In some cases, especially in dense deployments, abstraction of several nodes to provide a service may be beneficial. In order to enable such features, new protocols may have to be designed.

### 5.6 Link layer mesh routing

LoWPANs must support various topologies including mesh and star. Mesh topologies imply multi-hop routing. In this case, intermediate devices act as packet forwarders at the link layer. Typically these are FFDs that have more capabilities in terms of power, computation, storage and so on. This requires the routing protocol to affect minimal overhead on data packets independently of the number of hops, because of the reasons described in 5.1.

As stated in section 5.3, routing protocols should have low routing overhead (low chattiness), balanced with supporting topology changes and power conservation. The computation and memory requirements in the routing protocol should be minimal to satisfy the low cost and low power objectives. Thus, storage and maintenance of large routing tables is detrimental. Routing in presence of sleeping nodes must also be considered.

Routing issues are not within the scope of this paper, but we just want to mention two existing protocols and their issues. There is much published work on ad-hoc multi hop routing for devices. Some examples include [23], [24] and [25], all experimental. Also, these protocols are designed to use IP-based addresses that have large overheads. For example, the Ad hoc On-Demand Distance Vector (AODV) routing protocol uses 48 octets for a route request based on IPv6 addressing. Given the packet-size constraints, transmitting this packet without fragmentation and reassembly may be difficult. Thus, care should be taken when using existing routing protocols (or designing new ones) so that the routing packets fit within a single IEEE 802.15.4 frame. The Dynamic Source Routing protocol (DSR) is a simple and efficient routing protocol designed specifically for use in multi-hop wireless ad hoc networks of mobile nodes. DSR allows the network to be completely self-organizing and self-configuring, without the need for any existing network infrastructure or administration. The Ad hoc On Demand Distance Vector (AODV) routing algorithm is a routing protocol designed for ad hoc mobile networks. AODV is capable of both unicast and multicast routing. In general, AODV sends many small routing control packets, while DSR sends less, but bigger control packets.

### 5.7 IP Routing over a mesh of 802.15.4 nodes

When an IP protocol stack is implemented for LoWPAN devices, it should be possible to route based on the IP address over an 802.15.4 mesh. This solution must be carefully studied, especially taking into consideration that the Zigbee stack does not have an IP layer.

### 5.8 Security issues

IPv6 over LoWPANs will require confidentiality and integrity protection. This can be provided at the application, transport, network, and/or at the link layer. In all these cases, prevailing constraints will influence the choice of a particular protocol. Some of the more relevant constraints are small code size, low power operation, low complexity, and small bandwidth requirements. Some examples for threats that should be considered are man-in-the-middle attacks and denial of service attacks. A separate set of security considerations apply to bootstrapping a 6LoWPAN device into the network, for example for initial key establishment.

For network layer security, two models are applicable: end-to-end security (for example using IPSec transport mode), or security that is limited to the wireless portion of the network, (for example using a security gateway and IPSec tunnel mode). The disadvantage is the larger header size, which is significant at the LoWPAN frame MTUs.

## 6. Existing Approaches

So far we have shown the challenges that must be faced when thinking at developing a protocol that supports IPv6 over LoWPANs. We must always keep in mind that 802.15.4 and Zigbee are two completely different things, despite being related, so the possible approaches are completely different, due to the different nature of the two protocols. The 802.15.4 standard does not say anything about upper layers, whereas Zigbee specifications clearly adopt a network layer that acts as a routing layer as IP does, but this networking layer is, of course, different from IP. So, it's quite obvious that IPv6 cannot replace the Zigbee network layer.

In general, Sensor networks and IP can play together by means of two approaches: full IP stack throughout or edge network approach. In this chapter we analyze or mention some of the existing approaches for IPv6 over 802.15.4 and IPv6 over Zigbee.

6.1 Overview

6LoWPAN is a new standard that has been created by the Internet Engineering Task Force (IETF) that promises a true open network based on IPv6. It is openly available and provides an easy implemented model of Internet connectivity and interoperability with IP-based legacy systems, networks, applications, and tools; basically it is a simple alternative to Zigbee and other solutions. Arch Rock and Sensinode, two wireless network vendors, recently conducted successful interoperability demonstrations of the IETF 6LoWPAN standard. This solution basically applies an adaptation layer in between the MAC layer and the network layer (IPv6), in order to allow compatibility with the standard TCP/IP stack. The adaptation layer is needed to overcome all the issues described in section 5.3, and the adopted solutions will be described in detail in section 6.2.

The IETF 6LoWPAN approach is completely different from the Zigbee approach. Consequently, a device with multiple link-layer interfaces and that will act as a gateway will behave differently depending upon whether its devices adopt 6LoWPAN or Zigbee. We have in fact two different kinds of gateways: 6LoWPANs and Zigbee gateways. Their structures are shown in figures 4 and 5. The main difference is where the conversion is made: 6LoWPANs gateways make the conversion at IPv6 layer, Zigbee gateways have to go upper in the protocol stack, since device discovery and lots of important features are performed inside the Zigbee Application Layer.

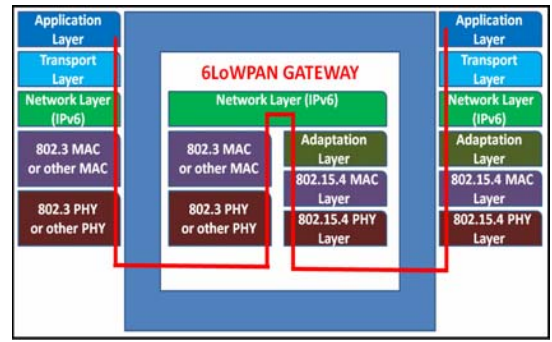

Fig. 4 6LoWPAN Gateway structure

A Zigbee Gateway is thus more complex than a 6LoWPAN gateway, and it is intended to provide an interface between Zigbee and IP devices through an abstracted interface on the IP side. The IP device is isolated from the Zigbee protocol by that interface: the Zigbee Gateway translates both addresses and commands between Zigbee and IP: the IP stack is thus terminated, and the gateway provides translation between the respective stacks acting as an agent on behalf of the IP device, isolating the IP device from the details of Zigbee operation and vice versa.

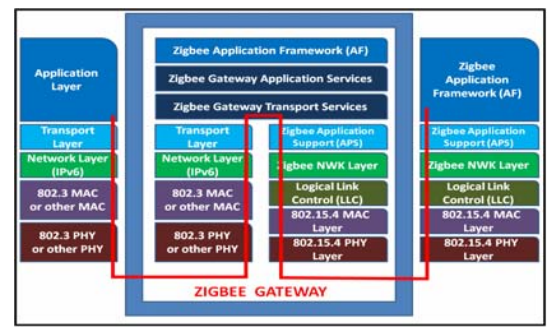

Fig. 5 Zigbee Gateway architecture

It must be remarked, that Zigbee Alliance has also developed the so called Zigbee Bridge or Zigbee Extension Device (ZED), that has a structure defined in figure 6, and extends the ZigBee network across an IP based network. A ZigBee Gateway Device (ZGD) defines instead a standard mechanism for accessing ZigBee networks from an IP network. The ZigBee network layer is continuous among the ZigBee devices by overlaying it on the IP network's transport layer. The ZED makes the IP connectivity transparent to the ZigBee devices. In an alternative configuration, a ZED may be used to communicate with IP devices that are executing the ZigBee stack and communicate through a ZigBee network layer. For example, the IP device can behave like an extension of the ZigBee network. The ZigBee stack runs over the IEEE 802.15.4 MAC and is encapsulated to run over the TCP/IP stack. The standardization of these devices will permit multiple vendors to interoperate and provide a superior solution to ZigBee users [18].

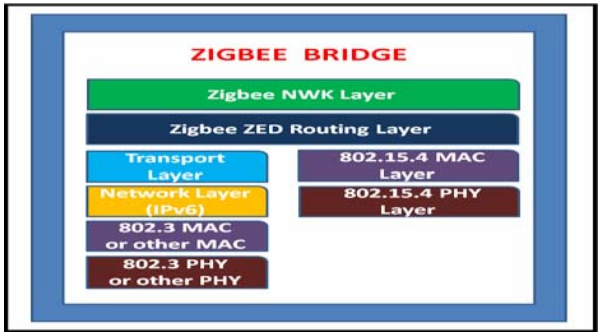

Fig. 6 Zigbee Bridge architecture

## 6.2 6LoWPAN

The IETF 6LoWPAN [10], [11], [12], [20] working group target was to define how to carry IP-based communication over IEEE 802.15.4 links in a manner to conform to open standards and to provide interoperability with other IP links and devices. In this way, it allows many different companies to manufacture LoWPAN devices that can work together in a network, and also allows these devices to work with the many networked computers and devices that already exist, thus eliminating the need for an array of complex gateways. Essentially, 6LoWPAN consists of an Adaptation Layer, that allows IEEE 802.15.4 frames to carry IPv6 on top of it, and solves the problems listed in section 5.3 by carrying the strictly necessary information in the frame, possibly compressed.

### 6.2.1 Compressed headers

A 6LoWPAN header includes, in order, of Mesh Addressing Header, Broadcast Header, Fragmentation Header, IPv6 Header, and UDP Header. In order to distinguish the different possible headers, the first byte is a so called Dispatch, it can assume the values shown in table 2, and it is divided into dispatch type (first two bits) and dispatch header (last six bits); its value indicates which headers follow and their ordering constraints relative to all other headers. The design decision for 6LoWPAN allows a very significant compression of the IPv6 header: instead of the original 40 bytes, an IPv6 header may be in some cases be represented using only 2 bytes. This is made possible by the introduction of the so called LOWPAN_HC1 compressed IPv6 header, that is 1 byte only, and it specifies if the fields that follows are compressed or not. In fact, IPv6 header fields cannot be always reduced to two bytes, although this situation would be very desirable. The fields that cannot be compressed will be carried in-line after the HC1_header. IPv6 header compression is possible by eliminating all those fields that can be retrieved somewhere else or that are fixed. For example the version field can be omitted since it is fixed, and both IPv6 source and destination addresses could be link local, so the IPv6 interface identifiers for the source or destination addresses can be inferred from the layer 2 addresses. In the same way the packet length can be derived from the Frame Length of the 802.15.4 PPDU (see figure 2), or from the datagram_size field in a fragment header. If there is no need for Traffic Class and Flow Label they will be assumed to have the fixed value zero, and the next header can be only UDP, ICMP or TCP. There is thus just one field that must be carried in full, and this field is 8 bits long: the Hop Limit. This compression can be applied to most IPv6 packets. If any field needs to be carried in-line, then the corresponding bits of the HC1_header will be set accordingly, and the fields will follow the HC1_header in this order: source address prefix and/or interface identifier, destination address prefix and/or interface identifier, Traffic Class, Flow Label and Next Header. The 8 bits of the HC1_header are thus enough to know what will follow the header: every bit represent if a specific field is uncompressed or elided. So, the first two bits will specify if the IPv6 source address is carried uncompressed later or if it is derived from link local prefix, bits 2 and 3 do the same with the IPv6 destination address, bit 4 is associated to Traffic Class and Flow Label and bits 5 to 7 show the next header and if the HC2 encoding format is used. The next header can be of 4 types: not compressed (00), UDP (01), ICMP (10) and TCP (11). HC2 encoding takes care of the possibility to also compress the transport header.

The UDP header itself may in fact be compressed: the corresponding HC2 encoding design concept is exactly the same as for HC1: we can compress source port, destination port, and length. As with HC1 there is one field that is carried full in mandatorily: the checksum. Length can be retrieved elsewhere, so it can be omitted (Payload Length from the IPv6 header minus the length of any extension headers present between the IPv6 header and the UDP header). Source and destination port can be in fact compressed to 4 bits and the added to the number 0xF0B0.

This approach allows reducing UDP header to 4 octets instead of the original 8. TCP and ICMP headers are not compressed.

Table 2: 6LoWPAN Gateway structure

| TYPE | HEADER | |
|---|---|---|
| 00 | xxxxxx | Not a LoWPAN frame |
| 01 | 000001 | Uncompressed IPv6 addresses |
| 01 | 000010 | LOWPAN_HC1 compressed IPv6 |
| 01 | 010000 | LOWPAN_BC0 broadcast |
| 01 | 111111 | Additional dispatch byte follows |
| 10 | xxxxxx | Mesh header |
| 11 | 000xxx | First fragmentation header |
| 11 | 100xxx | Subsequent fragmentation header |

### 6.2.2 Fragmentation issues

For the reasons explained in 5.3.2, fragmentation is needed. 6LoWPAN requires that all fragments of an IP packet carry the same "tag", which is assigned sequentially at source of fragmentation. In addition to the tag field, the size field is also present and encodes the entire size of the IP packet before link-layer fragmentation. To distinguish the subsequent packets (that might not arrive in order, but they must all come within 60 seconds), the offset fields (8 bits long) can be used: the field is present only in the second and subsequent link fragments and shall specify the offset of the fragment from the beginning of the payload datagram. The first octet of the datagram has an offset of zero.

### 6.2.3 Mesh addressing

Mesh addressing is made underneath the IPv6 layer, and it is defined by a dispatch header of '10' and it is composed by the originator address, the final destination address and the hops left. Originator and final destination address can be 16 or 64 bits (see section 3.2.4) as indicated by bits 2 and 3 of the dispatch header. The hops left are also stored in the dispatch header in bits 4-7 and they are decremented by each forwarding node before sending this packet towards its next hop.

### 6.2.4 Stateless address autoconfiguration and service discovery

The Interface Identifier [26] for an IEEE 802.15.4 interface may be based on the EUI-64 identifier assigned to the IEEE 802.15.4 device. In this case, the Interface Identifier is formed from the EUI-64 according to the "IPv6 over Ethernet" specification [27]. But since 802.15.4 devices can also have only 16-bit short addresses, a "pseudo 48-bit address" is formed as follows.

First, the left-most 32 bits are formed by concatenating 16 zero bits to the 16-bit PAN and they are concatenated with the 16-bit short address. The interface identifier is formed from this 48-bit address as per the "IPv6 over Ethernet" specification.

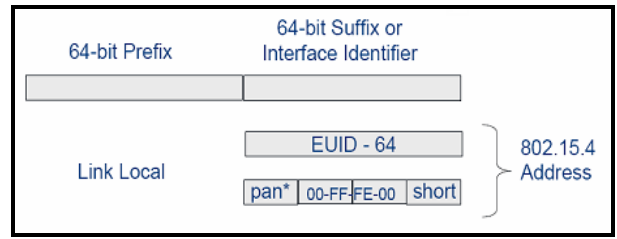

Fig. 7 IPv6 Address Assignment to Zigbee nodes

Service discovery for 6LoWPAN is performed via NDP (IPv6 neighbor discovery). This would expect mechanisms like link-layer multicast capabilities, but 802.15.4 links do not support link-layer multicast by default. One solution would be to use layer 2 broadcast functionality to distribute multicast packets in the network; however, an intensive use of broadcast would lead to a significant consumption of bandwidth, processing power, and battery power in sensor networks, and would lead to problems described in 5.3.3. Further information on investigation on NDP and SAA optimization for 802.15.4 sensor networks is provided by [19].

### 6.2.5 IP routing over a mesh of 802.15.4 nodes

This functionality comes for free with 6LoWPAN approach, since IP has always done multi-hop routing, and routers connect sub-networks to one another, and exchange messages using basic communication capabilities and protocols. They use routing tables to determine which node represents the "next-hop" toward the destination. Thus, IP routing over 6LoWPAN links does not require additional header information at 6LoWPAN layer.

### 6.2.6 Broadcast

Broadcast and Multicast functionalities can be used only in a mesh-enabled LoWPAN, and the approach behind them consists in a special dispatch value (the so called LOWPAN_BC0) followed by the sequence number, an 8 bits field, whose goal is to detect eventual packet duplication. The full specifications of capabilities are out of the scope of this document.

## 6.3 A translation method between 802.15.4 nodes and IPv6 nodes

Sakane et al. [13] propose a translation method which enables the internetworking between IPv6 and 802.15.4 node. The approach introduces a new device identifier (devid) to uniquely identify an IPv6 node or 802.15.4 node in a similar way which can be used by the application to transparently identify the peer node. Moreover they have also introduced the Dynamic Configuration 802.15.4 Aggregator (DCAGGR) for name resolution in 802.15.4 as there is no existing infrastructure for this purpose that can be used. To accommodate devids, they introduced an application header, which specifies both source and destination devids, at the top of the application payload. Every node at first registers to the translator (that is usually the PAN Coordinator) with their devid and corresponding IPv6 or ZigBee address. The proposed translation method assumes that there is a packet forwarding mechanism in 802.15.4 network, and the translator can resolve an 802.15.4 address assigned to the IPv6 node. When any 802.15.4 node want to communicate with an IP node it sends the packet to the translator which resolves the 802.15.4 address into the corresponding IPv6 address. After that, the packet is delivered to the destination IPv6 node by the delivery mechanism of IPv6 network.

The delivery of packet from IPv6 node to 802.15.4 is implemented in the same way, but the challenges are represented by the fact that the translator does not support fragmentation and reassembly. Thus, to deal with the heterogeneity of packet size while transmitting from IPv6 node to ZigBee node the application in the IPv6 network should be designed to prevent that an 802.15.4 node reassembles packets. The application in the 802.15.4 network should be also designed to prevent the node from fragmenting packets.

This approach does not work on cross regional areas as every node is registered with a preconfigured gateway.

## 6.4 Internetworking between Zigbee/802.15.4 and IPv6/802.3 networks

Wang, et al. [14] present a novel gateway design which can overlay the ZigBee/802.15.4 and the IPv6/802.3 networks together and allow internetworking among them. This architecture can easily be extended to all kind of IPv6 networks such as IPv6/802.11 or IPv6/UMTS. The goal of their approach is to interoperate ZigBee networks and IPv6 enabled networks such that a ZigBee node will be able to communicate with an IPv6 enabled node. To achieve this goal, five criteria have to be met. These design criteria was selected very carefully in such a way that they address the needs for internetworking between ZigBee and IPv6. The design criteria are the following:

- Each ZigBee node should be assigned a global unicast IPv6 address.
- Each IPv6 host which may communicate with a ZigBee node should obtain a ZigBee address.
- Service discovery should be propagated to different network domain.
- Broadcast data in ZigBee network should be transferred to the proper IPv6 hosts.
- Data packet transformations in the gateways should be as simple as possible and should not break the end-to-end model above the transport layer.

In this design, each ZigBee device is assigned a Global Unicast IPv6 address so that every IPv6 node can communicate with it directly. This IPv6 address assignment to the ZigBee node is using simple prefix delegation method. Here the gateway will support the functionality of prefix delegation and play its role as requesting router since ZigBee nodes cannot perform IPv6 Stateless Autoconfiguration, and since the nodes would suffer of memory problems if they have to keep IPv6 addresses. The address assignment is done simply adding delegated prefix with the 64 bit extended address of the ZigBee node hence mitigating the challenge mentioned in section 5.3.4. The gateway can easily remove the prefix part of the destination address and get the destination ZigBee address. The IPv6 address does not really exist on the ZigBee nodes. It is actually a pseudo address and kept in the gateway in a unicast mapping table. Moreover the Zigbee extended address is a unique address so the gateways need not to perform the DAD (Duplicated Address Detection) process. This fulfils first and fifth criteria.

On the other hand, each IPv6 node who wants to communicate with the ZigBee devices is also assigned with a ZigBee short address. UPnP (Universal Plug and Play) SSDP (Simple Service Discovery Protocol) helps to achieve this. They also solve the challenge described in section 5.3.5. When an IPv6 node wants to communicate with a Zigbee node, the first task is to find out the Zigbee coordinator. This is done by SSDP protocol with the PAN ID as the keyword. When a ZigBee coordinator gets a SSDP service it will transform the packet to ZigBee Service Discovery format and pass it to the Zigbee/802.15.4 network. The transformation will keep the record in the mapping table for a period so that the ZigBee response address assignment packet can reply to the proper IPv6 host. This procedure helps to fulfil the second criterion. Moreover the UPnP also helps to two way service discovering process and fulfilling third criterion. All the service discovery functions for Zigbee are defined in ZDO (Zigbee Device Object) and will be transformed to the XML format, which is necessary for SSDP and vice versa. The use of two ways service discovery process using UPnP helps to overcome the problem of preconfiguration with the gateway during bootstrap faced by the approach described in section 6.3. According to fourth criterion, Zigbee must support broadcasting as it is an ad hoc wireless network. To support this a IPv6 Multicast Group is also established keeping the gateway as the rendezvous point for relaying broadcast messages from ZigBee network to all correlated IPv6 nodes.

From the viewpoint of a ZigBee node, every IPv6 host is like another ZigBee node because it has a ZigBee address for communication and vice versa. The gateways will handle all the transformation like an IP-switching mechanism. All the transformation is done under layer 3 so there is no need to dig the application layer information which provides a way to keep the end to end security as well as to apply security on application layer to safeguard the data and mitigating challenge of security issue mentioned in section 5.3.8.

To deal with the frame size heterogeneity and fragmentation issue mentioned in section 5.3.1 and section 5.3.2 every IPv6 node sends data packet in a standard IPv6 packet, with the payload containing ZigBee APL data type which is used to communicate with ZDO or ZDP. The benefit of the IPv6 payload following the ZigBee specification is to keep the security and limit the packet length so that it will not be too large while in transformation. The gap between ZigBee APL data size (94 bytes) and IPv6 MTU in 802.3 (1280 bytes) is filled with zero so that the gateway can transform the payload with just simply discard the filler bits, replace the IPv6 Header with ZigBee NWK header and forward the packet. This architecture also support the cross region communication which was not supported by the address translation solution described in section 6.3. Figure 8 compares this approach with the one described in section 6.3, when two ZigBee nodes in two different ZigBee networks want to communicate with each other across the IPv6 network. Cross regional communication is supported in this approach since it uses UPnP as service discovery protocol, which makes the network more flexible. Moreover it does not bother to manage lots of pre-assigned parameters such as the gateway address at the time of bootstrapping.

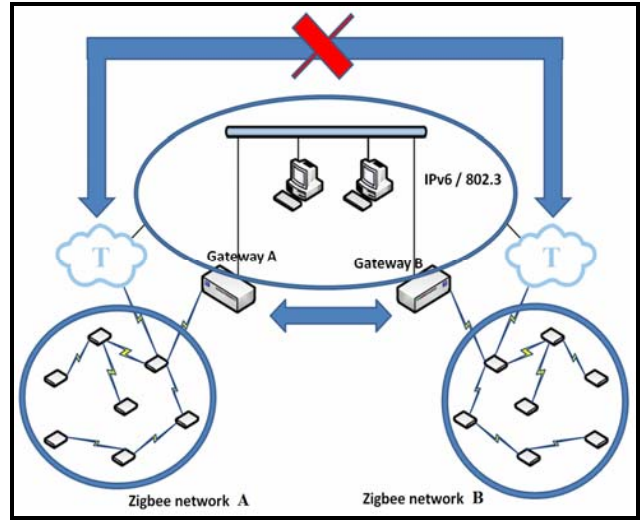

Fig. 8 Comparison between approach 6.3 and 6.4

### 6.5 Other existing approaches

The main approaches existing for internetworking solutions are the ones described above. Wang et al. [14] also mentions IP-NET as existing internetworking solution. IP-NET is a proprietary solution which is designed by the Helicomm Inc., a wireless solutions company with a strong market presence in Asia. There is unfortunately a lack of information on this solution, due to the proprietariness of the protocol. However, we know from [15] that a dual stack approach is adopted, in which both the 6LoWPAN and Zigbee stack coexist on the same 802.15.4 MAC, but in which they used once per time. Consequently, this is not an internetworking solution.

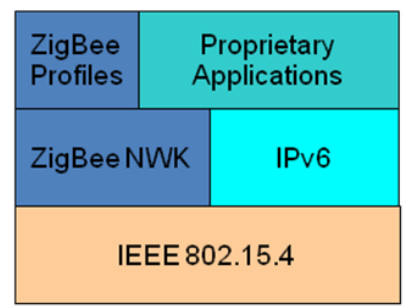

Fig. 9 IP-NET Stack

## 7. 6LoWPAN and Zigbee Interoperability Considerations

Both 6LoWPAN and ZigBee uses the IEEE 802.15.4 PHY/MAC and the major difference is whether or not IPv6 is enabled [17]. Zigbee has its own hierarchical routing algorithm and its own network addresses to communicate between FFDs and RFDs. 6LoWPAN is entirely leaning to the IPv6 function within IETF, on the other hand, ZigBee is a de-facto standard through the ZigBee Alliance and its network architecture over IEEE 802.15.4 MAC is composed of several layers except IP function. Given the IPv6 utility (described in chapter 5.2), 6LoWPAN is able to communicate using the unique IPv6 address. Interoperability between 6LoWPAN and ZigBee, however, is not feasible, and the main reasons why interoperability is compromised are:

- Header compressions described in 6.2.1 cannot be applied to Zigbee nodes;
- Service discovery is implemented on different layers;
- Some features of 802.15.4 MAC have been modified by 6LoWPAN for a more efficient usage, like association/disassociation, beacon and beaconless mode and many others.
- The three types of devices that Zigbee defines are unknown to 6LoWPAN.
- 6LoWPAN is composed of the traditional TCP/IP stack architecture, thus IPv6 stack is the transport layer beneath. In case of ZigBee, IPv6 will be one of the profiles of ZigBee.
- It does not matter for both to adopt IEEE 802.15.4 security algorithm, however ZigBee designed its own optimized security algorithm and that does not work on 6LoWPAN nodes. [20] analyzes 6LoWPAN security.

## 8. A Proposal for an IPv6 Interoperable Gateway

The existing internetworking approaches described in section 6 exploit different concept techniques and different layers. However, as shown in section 7, 6LoWPAN and Zigbee have proprietary solutions for IPv6 connection that are not compatible. A gateway that integrates both functionalities, i.e. that allows both a 6LoWPAN and Zigbee device indifferently to be converted to IPv6, would be desirable. Recently there is no movement towards such a solution. But if we apply the IP-NET concept of a dual stack approach on layer 2 in one side of the gateway, we can obtain an interoperable gateway as shown in figure 10. This is however only a proposal and still needs further studies and work on it.

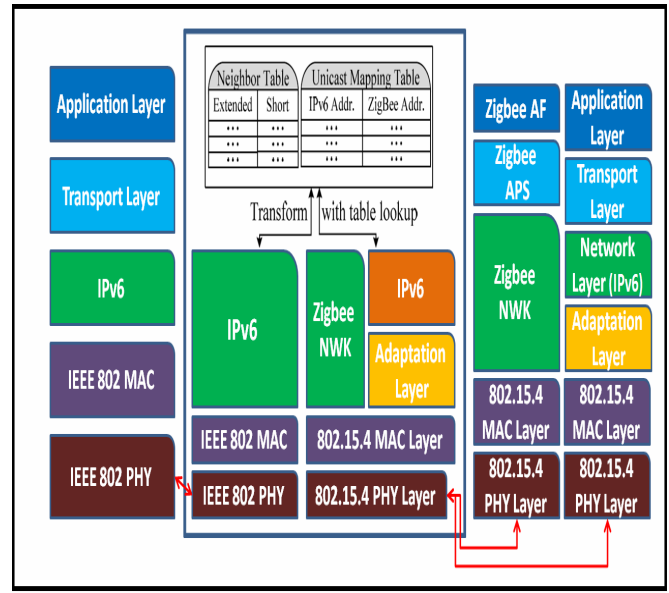

Fig. 10 Possible interoperable gateway between IPv6, 6LoWPAN and Zigbee

## 9. Conclusions

This paper deals with the interoperability of LoWPAN networks based on the IEEE 802.15.4 standard and mainly focuses on the implications and challenges of their connectivity to the Internet. We paid attention mostly on the current approaches of IP-LoWPAN internetworking solutions, giving more emphasis on ZigBee. There is no unique solution that mitigates the challenges that we listed out because of the heterogeneous network infrastructure and protocol stack. Future standardization of interconnecting devices like gateway or bridges may enable multiple LoPWAN networks to interoperate and provide internetworking solutions for global Internet. Researches, developers, and manufactures need to work together to explore this area in order to mitigate the challenges.

**Md. Sakhawat Hossen** is doing his MS in Internetworking in the Royal Institute of Technology (KTH), Sweden. He received his BSc degree in Computer Science and Information Technology from Islamic University of Technology (IUT), Gazipur, Bangladesh in 2004. Formerly he was a lecturer in the department of Computer Science and Engineering (CSE) of Stamford University Bangladesh. His research interest is mainly focused on Evolutionary optimization, Internet security, wireless sensor network (WSN), IP network and VoIP. Currently he is doing his Master thesis on secure session initiation protocol (SIP) user agent with key escrow capability to facilitate lawful interception (LI).

**A.F.M Sultanul Kabir** is currently working as a faculty member in American International University of Bangladesh (AIUB), Dhaka Bangladesh. Previously he worked as a faculty member in Shanto-Mariam University (SMUCT) during the period September 2009-December 2009. He also served as a lecturer in the department of CSE in University of Development Alternative (UODA) during the period January 2005- August 2006. During the period June 2008-May 2009 he worked as a researcher in Swedish Defence Research agency (FOI), Stockholm Sweden. He received his M.SC in Internetworking from Royal Institute of Technology (KTH), Stockholm Sweden in 2009. He also received his B.Sc degree in Computer Science and Information Technology from Islamic University of Technology (IUT), Gazipur, Bangladesh in 2004. His research interest mainly on Web 3.0 (Semantic web) and system security.

**Razib Hayat Khan** is doing his PhD at Department of Telematics, Norwegian University of Science and Technology (NTNU), Norway. He completed his M.Sc. in Information & Communication Systems Security specialized in Security in Open Distributed System from Royal Institute of Technology (KTH), Sweden in 2008. He worked under VRIEND project (http://vriend.ewi.utwente.nl) as part of his M.Sc. thesis which was sponsored by Sentinels, a joint initiative of the Dutch Ministry of Economic Affairs, the Netherlands organization for Scientific Research Governing Board and the Technology Foundation STW and the industrial partners were Philips Electronics, AkzoNobel, Corus, and DSM. He also worked as research engineer, Multimedia technologies at Ericsson AB, Sweden. He received his B.Sc. degree in Computer Science and Information Technology from Islamic University of Technology (IUT), Gazipur, Bangladesh in 2004. He served as a lecturer in Stamford University, Dhaka, Bangladesh during the period November 2004 – August 2006. He received the OIC (Organization of the Islamic Conference) scholarship for three years during his BSc studies. His research interest is mainly focused on Network performance modeling, Information Systems Security. At present he is working with performance and security issues in Communication system.

**Abdullah Azfar** is doing his MS in Erasmus Mundus NordSecMob program specialized in Security and Mobile Computing in Norwegian University of Science and Technology (NTNU), Norway and Royal Institute of Technology (KTH), Sweden. He received his BSc degree in Computer Science and Information Technology from Islamic University of Technology (IUT), Gazipur, Bangladesh in 2005. He served as a lecturer in the Islamic University of Technology during the period March 2006 – July 2008. He also served as a lecturer in Prime University, Dhaka, Bangladesh during the period October 2005 – February 2006. He received the Erasmus Mundus scholarship from the European Union for his MS studies and OIC (Organization of the Islamic Conference) scholarship for three years during his BSc studies. His research interest is mainly focused on Information Systems Security. At present he is working with security issues in VoIP.